\documentclass[twocolumn,pra,showpacs,superscriptaddress]{revtex4-1}

\usepackage{graphicx}
\usepackage[latin1]{inputenc}
\usepackage{graphicx,epstopdf}
\usepackage{dcolumn}
\usepackage{amsmath}
\usepackage[hypertex]{hyperref}
\usepackage{amsfonts}
\usepackage{amsthm}
\usepackage{amssymb}
\usepackage{mathrsfs}
\PassOptionsToPackage{caption=false}{subfig}
\usepackage{subfig}
\usepackage{color}
\usepackage{amsmath}
\usepackage{latexsym}

\newtheorem{theorem}{Theorem}

\newtheorem{proposition}[theorem]{Proposition}

\begin{document}

\title{One-norm geometric quantum discord under decoherence}

\author{J. D. Montealegre}
\email{jdmontealegrer@if.uff.br}

\author{F. M. Paula}
\email{fagner@if.uff.br}

\author{A. Saguia}
\email{amen@if.uff.br}

\author{M. S. Sarandy}
\email{msarandy@if.uff.br}

\affiliation{Instituto de F\'isica, Universidade Federal Fluminense, Av. Gal. Milton Tavares de Souza s/n, Gragoat\'a, 
24210-346, Niter\'oi, RJ, Brazil}

\date{\today}

\begin{abstract}
Geometric quantum discord is a well-defined measure of quantum correlation if Schatten 1-norm 
(trace norm) is adopted as a distance measure. Here, we analytically investigate the dynamical 
behavior of the 1-norm geometric quantum discord under the effect of decoherence. By starting 
from arbitrary Bell-diagonal mixed states under Markovian local noise, we provide the decays of 
the quantum correlation as a function of the decoherence parameters. In particular, we show that 
the 1-norm geometric discord exhibits the possibility of double sudden changes and freezing behavior 
during its evolution. For nontrivial Bell-diagonal states under simple Markovian channels, 
these are new features that are in contrast with the Schatten 2-norm (Hilbert-Schmidt) geometric discord. 
The necessary and sufficient conditions for double sudden changes as well as their exact locations in terms of 
decoherence probabilities are provided. Moreover, we illustrate our results by investigating 
decoherence in quantum spin chains in the thermodynamic limit.  

\end{abstract}


\maketitle

\section{Introduction}

Geometry plays a fundamental role in physics. In modern quantum theory, this idea has 
been perpetrated through many proposals of geometrization of quantum information 
resources, e.g., entanglement~\cite{Vedral:02,Shimony:95,Barnum:01,Wei:03}. In recent 
years, geometric approaches have also been extended for more general quantum correlation 
measures. In this scenario, special focus has been given to quantum discord (QD) (see 
Refs.~\cite{Modi,Celeri,Sarandy:12} for recent reviews), which is an information-theoretic 
measure of non-classical correlations originally proposed by Ollivier and Zurek~\cite{Ollivier} 
through an entropic (non-geometric) approach. 

The geometrization of QD, named as geometric quantum discord (GQD), has been proposed by Daki\'c, 
Vedral, and Brukner~\cite{Dakic}, mainly motivated by the difficulty of extracting analytical 
solutions for the entropic version of QD. GQD, as originally defined, can be seen as a deformation 
of the entropic QD~\cite{Costa:13} and quantifies the amount of 
quantum correlations in terms of its minimal Hilbert-Schmidt distance from the set of classical 
states. Its evaluation is analytically computable for general two-qubit states \cite{Dakic} as 
well as for arbitrary bipartite states \cite{Lu-Fu,Hassan,Rana:12}. Moreover, it has been shown 
to exhibit operational significance in specific quantum communication protocols (see, e.g., 
Refs.~\cite{Dakic:12,Gu:12}). Despite those remarkable features, GQD is known to be sensitive to 
the choice of distance measures (see, e.g., Ref.~\cite{Bellomo:12}). In turn, as recently pointed 
out~\cite{Hu:12,Tufarelli:12,Piani}, GQD as proposed in Ref.~\cite{Dakic} cannot be regarded as 
a good measure for the quantumness of correlations, since it may increase under local operations 
on the unmeasured subsystem. In particular, it has explicitly been shown by Piani in Ref.~\cite{Piani} 
that the simple introduction of a factorized local ancillary state on the unmeasured party changes 
the geometric discord by a factor given by the purity of the ancilla. From a technical 
point of view, the root of this drawback is the lack of contractivity of the Hilbert-Schmidt norm 
under trace-preserving quantum channels. Remarkably, this problem can be completely solved if 
Schatten 1-norm (trace norm) is adopted as a distance measure. Indeed, concerning general Schatten 
$p$-norms for a geometric definition of QD~\cite{Debarba:12}, it has been shown in Ref.~\cite{Paula:13} 
that the 1-norm is the only $p$-norm able to consistently define GQD (see also Ref.~\cite{Rana:13}).

In this work, we investigate the 1-norm GQD under the effect of decoherence. More specifically, we consider 
Bell-diagonal states going through Markovian local quantum channels that preserve the Bell-diagonal form. 
For these states, we analytically provide the 1-norm GQD as a function of the decoherence probability. 
In particular, we show that the 1-norm GQD may exhibit double sudden changes and freezing 
behavior during its evolution. Concerning double sudden changes, its occurrence for Bell-diagonal 
states under the simplest Markovian evolutions is a new feature that have not been found for both the 
entropic~\cite{Maziero:09} and the Schatten 2-norm geometric~\cite{Lu:10} QD. Specifically about the 
freezing behavior for general Bell-diagonal states under Markovian noise, it also occurs for the 
entropic QD~\cite{Mazzola:10}, being interpreted as a sudden transition between classical and quantum 
decoherence. We illustrate double sudden changes and freezing of the 1-norm GQD through simple examples. 
Moreover, we also consider the dynamical properties of quantum correlations in the XXZ quantum spin chain 
in the thermodynamic limit, showing a connection between the occurrence of sudden changes and the quantum 
phases of the XXZ model.

\section{GQD and Schatten 1-norm}

Let us begin by considering a bipartite system $AB$ in a Hilbert space ${\cal H}={\cal H}_A\otimes {\cal H}_B$. The system is 
characterized by quantum states described by density operators $\rho \in {\cal B}({\cal H})$, where 
${\cal B}({\cal H})$ is the set of bound, positive-semidefinite operators acting on ${\cal H}$ with 
trace given by ${\textrm{Tr}}\,\left[\rho\right]=1$. The 1-norm GQD between $A$ and $B$ is defined 
through the trace distance between $\rho$ and the closest classical-quantum state $\rho_{c}$, reading
\begin{equation}\label{eq:DG}
D_{G}(\rho)=\min_{\Omega_0}\left\Vert \rho-\rho_{c}\right\Vert_{1},
\end{equation}
where $\left\Vert X\right\Vert_{1}={\textrm{Tr}}\left[\sqrt{X^{\dagger}X}\right]$ is the $1$-norm 
(trace norm) and $\Omega_0$ is the set of classical-quantum states, whose general form is given by
\begin{equation}
\rho_{c}=\sum_{k}p_{k}\Pi^A_{k}\otimes\rho^{B}_{k} ,
\label{rho-classical}
\end{equation}
with $0 \le p_k \le 1$ ($\sum_k p_k = 1$), $\{\Pi^a_{k}\}$ denoting a set of orthogonal projectors 
for subsystem $A$, and $\rho^{B}_{k}$ a general reduced density operator for subsystem $B$. 
 
We will focus here in the particular case of two-qubit Bell diagonal states, whose density operator 
presents the form
\begin{equation}\label{bell}
\rho=\frac{1}{4}\left[I\otimes I+\vec{c}\cdot \left(\vec{\sigma}\otimes\vec{\sigma}\right)\right],
\end{equation}
where  $I$ is the identity matrix, $\vec{c}=\left(c_{1},c_{2},c_{3}\right)$ is a three-dimensional 
vector such that $-1\le c_i \le 1$ and $\vec{\sigma}=\left(\sigma_{1},\sigma_{2},\sigma_{3}\right)$ is a vector composed by 
Pauli matrices. In matrix notation, we have 
\begin{equation}
\rho=\frac{1}{4}\left( 
\begin{array}{cccc}
1+c_3   & 0        & 0        & c_1-c_2 \\ 
0       & 1-c_3    & c_1+c_2  & 0 \\ 
0       & c_1+c_2  & 1-c_3    & 0 \\ 
c_1-c_2 & 0        & 0        & 1+c_3
\end{array}
\right) .  \label{rho-bell-mat}
\end{equation}%
The minimization over the whole set of classical states was obtained for GQD as defined by the 
2-norm~\cite{Lu-Fu} and by the relative entropy~\cite{Modi:10}. For those cases, it can be proved 
that the minimal state is a decohered (measured) state, i.e. the minimization runs over 
classical states of the form $\rho_c = \Phi (\rho)$, where
\begin{equation}
\Phi(\rho)=\sum_{k=-,+}\left(\Pi_{k}\otimes I\right)\rho \left(\Pi_{k}\otimes I\right),
\end{equation}
with
\begin{equation}
\Pi_{\pm}=\dfrac{1}{2}\left(I\pm \vec{n}\cdot\vec{\sigma}\right)
\end{equation}
and $\vec{n}=\left(n_{1},\, n_{2},\,n_{3}\right)$ is an unitary vector. For the 1-norm GQD, minimization 
over decohered states has also been shown to be sufficient for the case of Bell-diagonal states~\cite{Paula:13}. 
In this case, the 1-norm GQD can be analytically computed, yielding~\cite{Paula:13}
\begin{equation}
D_{G}(\rho) = \text{int}\left[\left|c_{1}\right|,\left|c_{2}\right|,\left|c_{3}\right|\right], 
\label{1normGD}
\end{equation}
where $\text{int}\left[\left|c_{1}\right|,\left|c_{2}\right|,\left|c_{3}\right|\right]$ describes 
the {\it intermediate} result among the absolute values of the correlation functions $c_{1}$, $c_{2}$ and $c_{3}$. 
More specifically, for $\left|c_{i}\right| \le \left|c_{j}\right| \le \left|c_{k}\right|$ ($i,j,k \in \{1,2,3\}$), we 
attribute $\left|c_{j}\right| = \text{int}\left[\left|c_{i}\right|,\left|c_{j}\right|,\left|c_{k}\right|\right]$.  
Remarkably, this is equivalent to the {\it negativity of quantumness}, which is a measure of nonclassicality 
recently introduced in Refs.~\cite{Piani:NoQ,Nakano:12} and experimentally discussed  in Ref.~\cite{Silva:12}. 
Then, for Bell-diagonal states, by investigating the general behavior of the 1-norm GQD under decoherence, 
we are automatically providing the behavior of the negativity of quantumness under decoherence.

\section{1-norm GQD under decoherence}

We will consider the system-environment interaction through the operator-sum representation 
formalism~\cite{Nielsen-Chuang}. In this context, we will take the evolution of a quantum state 
$\rho$ as described by a trace-preserving quantum operation $\varepsilon(\rho)$, which is given by 
\begin{equation}
\varepsilon(\rho) = \sum_{i,j} \left(E_i\otimes E_j\right) \rho \left(E_i \otimes E_j\right)^\dagger, 
\end{equation}
where $\{E_k\}$ is the set of Kraus operators associated to a decohering process of a single qubit, 
with the trace-preserving condition reading $\sum_k E_k^\dagger E_k = I$. We provide in Table~\ref{t1} 
a list of Kraus operators for a variety of channels considered in this work.

\begin{table}[hbt]
\begin{tabular}{|c|c|}
\hline
 & $\textrm{Kraus operators}$                                         \\ \hline \hline
 & \\ 
BF   & $E_0 = \sqrt{1-p/2}\, I , E_1 = \sqrt{p/2} \,\sigma_1$                        \\ \hline
 & \\ 
PF   & $E_0 = \sqrt{1-p/2}\, I , E_1 = \sqrt{p/2}\, \sigma_3$                        \\ \hline
 & \\ 
BPF & $E_0 = \sqrt{1-p/2}\, I , E_1 = \sqrt{p/2} \,\sigma_2$                        \\ \hline
 & \\ 
GAD   & 
$E_0=\sqrt{p}\left( 
\begin{array}{cc}
1 & 0 \\ 
0 & \sqrt{1-\gamma} \\ 
\end{array} \right) , 
E_2=\sqrt{1-p}\left( 
\begin{array}{cc}
\sqrt{1-\gamma} & 0 \\ 
0 & 1 \\ 
\end{array} \right)$  \\
& \\
 & $E_1=\sqrt{p}\left( 
\begin{array}{cc}
0 & \sqrt{\gamma} \\ 
0 & 0 \\ 
\end{array} \right) ,
E_3=\sqrt{1-p}\left( 
\begin{array}{cc}
0 & 0 \\ 
\sqrt{\gamma} & 0 \\ 
\end{array} \right)$  \\ \hline
\end{tabular}
\caption[table1]{Kraus operators for the quantum channels: bit flip (BF), phase flip (PF), 
bit-phase flip (BPF), and generalized amplitude damping (GAD), where 
$p$ and $\gamma$ are decoherence probabilities. }
\label{t1}
\end{table}

The decoherence processes BF, PF, and BPF in Table~\ref{t1} preserve the Bell-diagonal form of the 
density operator $\rho$. For the case of GAD, the Bell-diagonal form is kept for arbitrary $\gamma$ 
and $p=1/2$. In this situation, we can write the quantum operation $\varepsilon(\rho)$ as
\begin{equation}
\varepsilon(\rho)=\frac{1}{4}\left( 
\begin{array}{cccc}
1+c^\prime_3   & 0        & 0        & c^\prime_1-c^\prime_2 \\ 
0       & 1-c^\prime_3    & c^\prime_1+c^\prime_2  & 0 \\ 
0       & c^\prime_1+c^\prime_2  & 1-c^\prime_3    & 0 \\ 
c^\prime_1-c^\prime_2 & 0        & 0        & 1+c^\prime_3
\end{array}
\right) ,  \label{eps-rho-bell-mat}
\end{equation}%
where the values of the correlation vector ${\vec{c^\prime}} = (c^\prime_1$, $c^\prime_2$, $c^\prime_3)$ 
are given in Table~\ref{t2}.
\begin{table}[hbt]
\begin{tabular}{|c|c|c|c|}
\hline
$\textrm{Channel}$ & $c^\prime_1$      & $c^\prime_2$     & $c^\prime_3$      \\ \hline \hline
& & & \\ 
BF                 &  $c_1$            & $c_2 (1-p)^2$    & $c_3 (1-p)^2$     \\ \hline
& & & \\ 
PF                 &  $c_1 (1-p)^2$    & $c_2 (1-p)^2$    & $c_3$             \\ \hline
& & & \\ 
BPF                &  $c_1 (1-p)^2$    & $c_2$            & $c_3 (1-p)^2$     \\ \hline
& & & \\ 
GAD                &  $c_1 (1-\gamma)$ & $c_2 (1-\gamma)$ & $c_3 (1-\gamma)^2$ \\ \hline
\end{tabular}
\caption[table2]{Correlation functions for the quantum operations: bit flip (BF), phase flip (PF), 
bit-phase flip (BPF), and generalized amplitude damping (GAD). For GAD, we  
fixed $p=1/2$.}
\label{t2}
\end{table}

Since the Bell-diagonal form is preserved in Eq.~(\ref{eps-rho-bell-mat}), we can directly obtain 
the 1-norm GQD from Table~\ref{t2} by using that
\begin{equation}
D_{G}[\varepsilon(\rho)] = \text{int}\left[\left|c^\prime_{1}\right|,\left|c^\prime_{2}\right|,
\left|c^\prime_{3}\right|\right]. 
\label{1normGD-eps}
\end{equation}
Then, we can obtain a general pattern of the 1-norm GQD decay as a function of the decoherence parameters. 
In particular, depending on both the original density operator $\rho$ and the quantum channel, we may have 
sudden changes in the decay rate of $D_G(\rho)$. This will occur whether there are crossings among the components 
$\left|c^\prime_{1}\right|$, $\left|c^\prime_{2}\right|$, and $\left|c^\prime_{3}\right|$ as 
a function of $p$ or $\gamma$. Moreover, additional effects may appear, which corresponds to  
{\it double sudden changes} and {\it freezing behavior}. More specifically, we can identify the following 
general behavior for the dynamics in Table~\ref{t2}: 

(i) BF, PF, and BPF: Double sudden changes occur if and only if the constant component $|c_k|$ for the 
channel is nonvanishing and is the lowest absolute value of the components of the correlation vector ${\vec c}$, 
with the two other components exhibiting distinct values ($|c_k|=|c_1|$ for BF, $|c_k|=|c_3|$ for PF, and 
$|c_k|=|c_2|$ for BPF). For instance, take a density operator $\rho$ described by a correlation vector 
$\vec{c}$ such that $|c_1|>|c_2|>|c_3|\ne 0$. By passing $\rho$ through the PF channel, we will 
initially have $D_G[\varepsilon(\rho)] = |c_2^\prime|=|c_2|(1-p)^2$. Then, there will be a first crossing 
involving $|c_2^\prime|$ and the constant component $|c_3^\prime|=|c_3|$ for a critical point $p=p_1^{SC}$, 
which will imply $D_G[\varepsilon(\rho)] = |c_3|$. The 1-norm GQD will be {\it frozen} as a function of $p$ 
for a certain interval (robust to decoherence in this interval) and finally a new crossing will occur 
involving now $|c_3^\prime|$ and $|c_1^\prime|$ for a second critical point $p=p_2^{SC}$. From this point 
on, $D_G[\varepsilon(\rho)] = |c_1^\prime|=|c_1|(1-p)^2$ until $p=1$. Similar examples can be obtained for 
BF and BPF by replacing $c_3$ for $c_1$ and $c_2$, respectively. We also observe here that, in presence 
of symmetry, the double sudden change phenomenon may get degenerate into a single sudden change. As a 
simple example, this will happen in case of $U(1)$-symmetric states, for which $c_1=c_2$. By passing 
such a state through the PF channel, we will have at most a single sudden change (if $|c_3| < |c_1|=|c_2|$). 

(ii) GAD: Double sudden changes occur if and only if $|c_3|\,>\,|c_1|,|c_2|$ and $|c_1| \ne |c_2| \ne 0$. 
In this case, by passing $\rho$ through the GAD channel, we will initially have 
$D_G[\varepsilon(\rho)] = \textrm{Max}[|c_1^{\prime}|,|c_2^{\prime}|]=\textrm{Max}[|c_1|,|c_2|](1-\gamma)$. 
Then, there will be a first crossing involving $\textrm{Max}[|c_1^{\prime}|,|c_2^{\prime}|]$ and the 
component $|c_3^\prime|$ for a critical point $\gamma=p_1^{SC}$, which will imply 
$D_G[\varepsilon(\rho)] = |c_3^\prime| = |c_3| (1-\gamma)^2$. Finally, a new crossing will occur 
involving now $|c_3^\prime|$ and $\textrm{Min}[|c_1^{\prime}|,|c_2^{\prime}|]$ for a second critical point 
$\gamma=p_2^{SC}$. From this point on, $D_G[\varepsilon(\rho)] = \textrm{Min}[|c_1^\prime|,|c_2^\prime|] =%
\textrm{Min}[|c_1|,|c_2|](1-\gamma)$ until $\gamma=1$. Note that no freezing occurs here. We also observe that 
$U(1)$-symmetric states also constrain the dynamics to exhibit at most a single sudden change 
(if $|c_3| > |c_1|=|c_2|$). 

Hence, the 1-norm GQD exhibits two new aspects for nontrivial Bell-diagonal states [$|c_i| \ne 0$ ($\forall i$)] 
under Markovian evolution in comparison with the 2-norm GQD: (a) double sudden change as a function of its 
decay rate; (b) freezing behavior -- and therefore robustness against decoherence -- for a finite interval as 
a function of $p$ for the channels BF, PF, and BPF. These properties can be obtained for the 2-norm GQD 
only if either more complex states are considered (e.g., X states) or the system is subjected to non-Markovian 
dynamics~\cite{Karpat:11,Song:11,Song:12,Yuan:13,Franco:13}. Indeed, the impossibility of the 2-norm GQD to reveal 
double sudden change and freezing for Bell-diagonal states under the Markovian local channels considered 
in this work can be proven below.

\begin{proposition}
{\it Consider a nontrivial two-qubit Bell-diagonal state [ $|c_i| \ne 0$ ($\forall i$) ] evolving under  
Bell-diagonal-preserving Markovian local noise described by the BF, BPF, PF, or GAD channels. For such individual 
dynamical evolutions, the 2-norm GQD is unable to exhibit either double sudden change or nonvanishing freezing 
as a function of the decoherence parameters.} 
\begin{proof}
From Ref.~\cite{Lu:10}, we can write the 2-norm GQD as $D_2 = (1/4)(\sum_i |c_i^\prime|^2 - \max |c_i^\prime|^2)$. 
Then, in order for a double sudden change to occur, we must have two changes in the maximum value of 
$|c_i^\prime|^2$ as a function of the decoherence parameter, namely, $p$ for BF, PBF, PF, and $\gamma$ for GAD. 
This is because $\sum_i |c_i^\prime|^2$ is a smooth function of $p$ or $\gamma$. However, from Table~\ref{t2}, 
it can be observed that, for any channel, two of the correlation functions $|c_i^\prime|^2$ display the same decay 
rate, which means that they do not cross as a function of $p$ or $\gamma$. Therefore, at most one sudden change
is allowed for the 2-norm GQD. Concerning the freezing behavior, we must have $D_2$ as a constant for a fixed 
interval of $p$ or $\gamma$. By rewriting $D_2$ as $D_2 = (1/4)(\min |c_i^\prime|^2 + {\textrm{int}} |c_i^\prime|^2)$, 
it follows from Table~\ref{t2} that $D_2$ cannot be a constant. Indeed, as 
discussed before, two of the correlation functions $c_i^\prime$ depend on $p$ with the same decay rate. Moreover, 
the possibility of  two nonvanishing constant values for $c_i^\prime$ has been excluded. Hence, 
there are no double sudden changes or freezing of correlations for the 2-norm GQD in the case of nontrivial 
Bell-diagonal states [$|c_i| \ne 0$ ($\forall i$)] and for the dynamical evolutions in Table~\ref{t2}.
\end{proof}
\end{proposition}
Notice that, by relaxing the requirement  $|c_i| \ne 0$ ($\forall i$), we can obtain the 
freezing behavior for the 2-norm GQD $D_2$ for {\it{separable}} Bell-diagonal states 
(see, e.g., Ref.~\cite{Yao:12}). In particular, from 
$D_2 = (1/4)(\min |c_i^\prime|^2 + {\textrm{int}} |c_i^\prime|^2)$, the freezing would 
occur for the BF, PF, and PBF channels by taking the constant component $|c_k|$ as the 
intermediate absolute value of the components of ${\vec c}$ and $\min |c_i|=0$ ($i \ne k$). 
However, this is a very restrictive solution, which is monotonically related to the 1-norm GQD, 
i.e., $D_2=D_G^2/4$. In other words, the 1-norm GQD allows for the freezing behavior in a much 
wider class of Bell-diagonal states. A similar freezing behavior is also known for the entropic QD, 
where an interpretation of sudden transition between classical and quantum decoherence has been 
introduced~\cite{Mazzola:10}. The sudden change critical points $p=p_1^{SC}$ and $p=p_2^{SC}$ 
for the 1-norm GQD can be analytically obtained by imposing the equality of the absolute 
values of the components of $\vec{c^\prime}$ at the crossings described in the discussion 
above. The results are organized in Table~\ref{t3}. 
\begin{table}[hbt]
\begin{tabular}{|c|c|c|c|}
\hline
 &  &  &  \\ 
                   & $p_1^{SC}$                                         & $p_2^{SC}$     & Conditions      \\ \hline \hline
 & & & $|c_1| < |c_2|,|c_3|$ \\ 
BF                 &  $1-\sqrt{\frac{|c_1|}{\textrm{Min}(|c_2|,|c_3|)}}$  & $1-\sqrt{\frac{|c_1|}{\textrm{Max}(|c_2|,|c_3|)}}$ & 
$|c_1| \ne 0$ \\ & & & $|c_2| \ne |c_3|$       \\ \hline
  & & & $|c_3| < |c_1|,|c_2|$ \\ 
PF                 &  $1-\sqrt{\frac{|c_3|}{\textrm{Min}(|c_1|,|c_2|)}}$  & $1-\sqrt{\frac{|c_3|}{\textrm{Max}(|c_1|,|c_2|)}}$ & 
$|c_3| \ne 0$ \\ & & &  $|c_1| \ne |c_2|$    \\ \hline
  & & & $|c_2| < |c_1|,|c_3|$ \\ 
BPF                &  $1-\sqrt{\frac{|c_2|}{\textrm{Min}(|c_1|,|c_3|)}}$  & $1-\sqrt{\frac{|c_2|}{\textrm{Max}(|c_1|,|c_3|)}}$ & 
$|c_2| \ne 0$ \\ & & &  $|c_1| \ne |c_3|$    \\ \hline
  & & & $|c_3| > |c_1|,|c_2|$ \\ 
GAD                &  $1-\frac{\textrm{Max}(|c_1|,|c_2|)}{|c_3|}$  & $1-\frac{\textrm{Min}(|c_1|,|c_2|)}{|c_3|}$ & 
$|c_1|,|c_2| \ne 0$ \\ & & &  $|c_1| \ne |c_2|$   \\ \hline
\end{tabular}
\caption[table3]{Critical points $p_1^{SC}$ and $p_2^{SC}$ in terms of the components of the correlation vector 
$\vec{c}$. The conditions provided are necessary and sufficient for assuring double sudden changes. Notice that 
double sudden changes may get degenerate if symmetries (implying in the equality of some of the components of 
$\vec{c}$) are present in the quantum state.}
\label{t3}
\end{table}

\section{Examples}

\subsection{Double sudden change for the GAD channel}

\begin{figure}[ht!]
\includegraphics[scale=0.6]{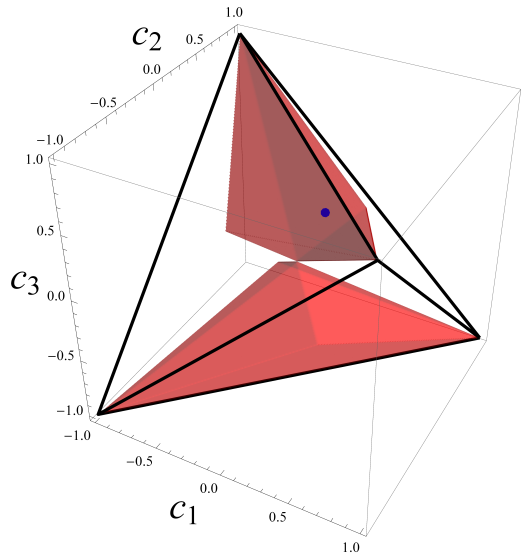} \caption{\label{f1} (Color online) Bell-diagonal states exhibiting double 
sudden changes under the GAD channel (red region inside the tetrahedron). The blue point represents the state 
${\vec{c}} = (0.1,0.2,0.3)$ used as the initial state in Fig.~\ref{f2}. }
\end{figure}

\begin{figure}[ht!]
\includegraphics[scale=0.82]{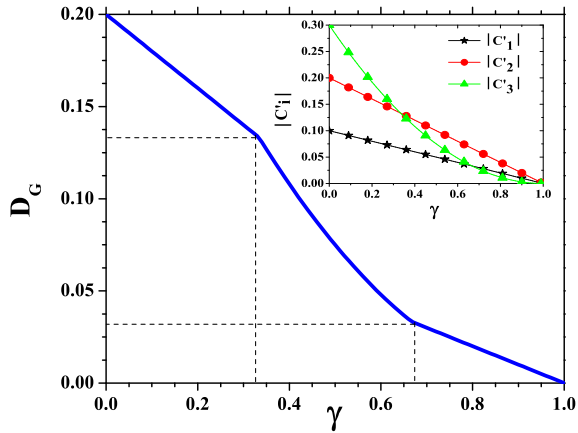} \caption{\label{f2} (Color online) Double sudden change for the state given 
by ${\vec{c}} = (0.1,0.2,0.3)$ under the GAD channel, 
with $p_1^{SC}=1/3$ and $p_2^{SC}=2/3$. Inset: Absolute value of the correlation functions as a function of $\gamma$.}
\end{figure}

We begin by illustrating the double sudden change behavior for the GAD channel. Let us take a quantum 
system described by a Bell-diagonal state $\rho$. In order for $\rho$ to correspond to a physical state, 
we must impose $\lambda_{i} \geq 0$ and $\sum_i \lambda_{i}=1$, with $\lambda_i$ denoting the eigenvalues of $\rho$. 
Under these conditions, the vector  $\vec{c}$ turns out to be restricted to a tetrahedron whose vertices situated 
on the points $(1,1,-1)$, $(-1,-1,-1)$, $(1,-1,1)$, and $(-1,1,1)$, as plotted in Fig.~\ref{f1}. Then, by using 
the conditions obtained in Table~\ref{t3} for the GAD channel, we can determine a physical region in the state 
space associated with double sudden changes in the decay rates of the 1-norm GQD. This is plotted in red color 
in Fig.~\ref{f1}. Within this physical region, we identify an explicit example to exhibit the double sudden 
change phenomenon, which is given by the values $c_1=0.1$, $c_2=0.2$, and $c_3=0.3$ (blue dot in Fig.~\ref{f1}). 
Its 1-norm GQD is then plotted in Fig.~\ref{f2}. From this plot, we observe the two critical points at $p_1^{SC}=1/3$ 
and $p_2^{SC}=2/3$, which are in agreement with the values in Table~\ref{t3}.

\subsection{Freezing behavior for the PF channel}

\begin{figure}[ht!]
\includegraphics[scale=0.82]{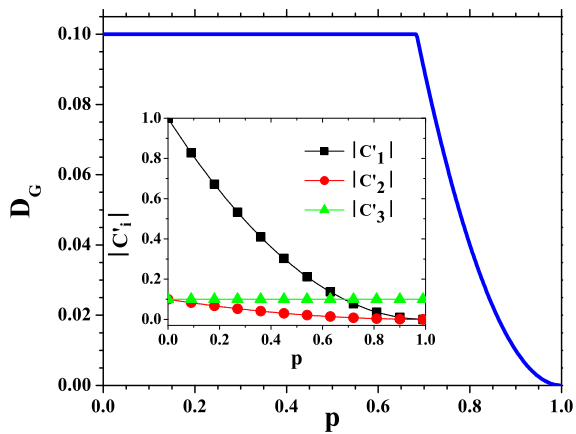} \caption{\label{f3} (Color online) Freezing behavior for the state given 
by ${\vec{c}} = (1,-0.1,0.1)$ under the PF channel. Inset: Absolute value of 
the correlation functions as a function of $p$. }
\end{figure}

Let us now illustrate the freezing behavior by using the PF channel. In this direction, we choose the state 
$c_1=1$ and $c_2=-c_3$ (with $c_3 \ne 0$), which has been discussed in Ref.~\cite{Mazzola:10} for the entropic 
QD. For this state, it is possible to show that the freezing behavior is followed by a sudden change at 
$p^{SC}=1-\sqrt{|c_3|}$ for both the entropic QD and the 1-norm GQD. Notice that the 2-norm GQD does not exhibit 
freezing for this state, since $|c_i| \ne 0$ ($\forall i$). Then, for the entropic QD and the 1-norm GQD, we 
obtain a large and controllable freezing interval. Indeed, by reducing $|c_3|$ (and consequently $D_G$), we 
can increase the freezing interval as close as desired to $p=1$. Naturally, the longer the freezing interval 
is, the less 1-norm GQD is available. This is illustrated in Fig.~\ref{f3}, where we have chosen $c_3=0.1$. 
We observe that a large freezing is obtained, with the sudden change occurring for $p>0.6$.

\subsection{Quantum spin chains under decoherence}

We consider now the correlation properties of a multi-qubit system under decoherence. More specifically, we will 
take a critical quantum spin chain, whose ground state is described by a Bell-diagonal state. In this context, 
we will illustrate the discussion of the 1-norm GQD between two spins by investigating the XXZ spin chain, 
whose Hamiltonian is given by
\begin{equation}
H_{XXZ}=-\frac{J}{2} \sum_{i=1}^{L} \left( \sigma^x_i \sigma^x_{i+1} + 
\sigma^y_i \sigma^y_{i+1} + \Delta \sigma^z_i \sigma^z_{i+1} \right),
\label{HXXZ}
\end{equation}
where periodic boundary conditions are assumed, ensuring therefore translation symmetry. 
We will set the energy scale such that $J=1$ and will be interested in a nearest-neighbor spin pair at 
sites $i$ and $i+1$. The model exhibits a first-order quantum phase transition at $\Delta=1$ and an 
infinite-order quantum phase transition at $\Delta=-1$. 
The entropic quantum discord has already been analyzed for nearest-neighbors, 
with the characterization of its quantum critical properties discussed~\cite{Marcelo,Dill:08}. 
For the geometric versions of QD in the XXZ model, see Ref.~\cite{Paula:13}. 
Concerning its symmetries, the XXZ chain exhibits $U(1)$ invariance, namely, 
$\left[H,\sum_i \sigma_z^i\right]=0$, which ensures that the two-spin density operator, for the chain in 
its ground state, is given by Eq.~(\ref{rho-bell-mat}). By using the Hellmann-Feynman 
theorem~\cite{Hellmann:37,Feynman:39} for the XXZ  Hamiltonian~(\ref{HXXZ}), we then obtain
\begin{eqnarray}
c_1 &=& c_2 = \frac{1}{2} \left(G_{xx} + G_{yy}\right) = \Delta \frac{\partial \varepsilon_{xxz}}{\partial \Delta} 
- \varepsilon_{xxz} \, , \nonumber \\
c_3 &=& G_{zz} = -2 \frac{\partial \varepsilon_{xxz}}{\partial \Delta} \, ,
\label{c-xxz}
\end{eqnarray}
where $\varepsilon_{xxz}$ is the ground state energy density 
\begin{equation}
\varepsilon_{xxz} = \frac{\langle \psi_0| H_{XXZ} |\psi_0 \rangle}{L} = - \frac{1}{2} \left(G_{xx} + 
G_{yy} + \Delta G_{zz} \right),
\label{aux-xxz}
\end{equation}
with $|\psi_0\rangle$ denoting the ground state of $H_{XXZ}$. Eqs.~(\ref{c-xxz}) and~(\ref{aux-xxz}) hold 
for a chain with an arbitrary number of sites, allowing the discussion of correlations either for finite 
or infinite chains. Indeed, ground state energy as well as its derivatives can be exactly determined by 
Bethe Ansatz technique for a chain in the thermodynamic limit~\cite{Yang:66}. Then, from Tables~\ref{t2} and 
\ref{t3}, we can obtain a general picture of the decoherence effects for different quantum phases. In particular, 
it is possible to show (see, e.g., Ref.~\cite{Marcelo}) that $c_1<c_3$ for both $\Delta<-1$ and $\Delta>1$. 
On the other hand, $c_1>c_3$ for the gapless disordered phase $-1<\Delta<1$. This implies, for the XXZ model, 
that the occurrence of sudden changes may be driven by quantum phase transitions, whose behavior is 
summarized in Table~\ref{t4}. 

\begin{table}[hbt]
\begin{tabular}{|c|c|c|c|}
\hline
$\textrm{Channel}$ & $\Delta<-1$       & $-1<\Delta<1$  & $\Delta>1 $             \\ \hline \hline
BF                 &  1 S.C.           & No S.C.        & No S.C.    \\ \hline
PF                 &  No S.C.          & No S.C.        & No S.C.                 \\ \hline
BPF                &  1 S.C.           & No S.C.        & No S.C.    \\ \hline
GAD                &  1 S.C.           & No S.C.        & No S.C.    \\ \hline
\end{tabular}
\caption[table2]{Occurrence of sudden changes (denoted here as S.C.) in the 1-norm GQD for different 
quantum channels. The decoherence behavior may change for different quantum phases. The phase $\Delta>1$ 
does not exhibit S.C. because $c_1=c_2=0$.}
\label{t4}
\end{table}

\begin{figure}[ht!]
\includegraphics[scale=0.82]{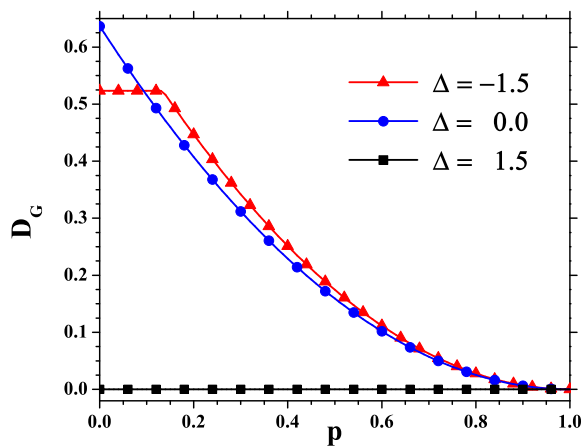} \caption{\label{f4} (Color online) Sudden change behavior for 
first neighbors in the ground state of the XXZ chain under the BF channel. Distinct phases may exhibit 
different behaviors, going from monotonic decay ($\Delta>-1$) to sudden change dynamics ($\Delta<-1$).}
\end{figure}

For a specific example, we consider the decoherence behavior for the BF flip channel in the three 
quantum regimes governed by the anisotropy $\Delta$. This is plotted in Fig.~\ref{f4}. As we can 
see, the sudden change behavior that is present in the antiferromagnetic phase ($\Delta=-1.5$ in the plot) 
disappears in the gapless phase ($\Delta=0$ in the plot), characterizing the infinite-order quantum 
phase transition of the XXZ chain. Therefore, decoherence turns out to be a probe for the antiferromagnetic 
quantum critical point. Such a characterization is a particular feature of the XXZ chain, 
holding possibly also for related models, but not implying a general approach for quantum criticality. 
For the entropic QD, a detection of quantum phase transitions in terms of sudden change behavior has 
recently been introduced in Refs.~\cite{Bose:12,Bose:13}.

\section{Conclusions}

We have analytically investigated the 1-norm GQD under the effect of Markovian noise for Bell-diagonal states. 
In particular, we have shown that, for the simplest cases of Markovian dynamics, the 1-norm GQD may already exhibit 
double sudden changes or freezing behavior. The origin of such features is the robustness of QD under decoherence. 
This inherent robustness can be traced back as a consequence of the negligibility of the set of classical states 
(zero discord states) in comparison with the whole state space~\cite{Ferraro:10}. This is in strong contrast with the 
set of separable (nonentangled) states, which exhibits finite volume in state space. This implies, e.g. in the 
possibility of sudden death for entanglement, while just sudden changes (typically with no death) in the case of QD. 
Specifically concerning the 1-norm GQD, it inherits this robustness of QD against noise and reveals it for simple 
Bell-diagonal states, which is a by-product of the trace norm properties. 
Experimental investigations of single sudden changes and freezing 
behavior have been realized through different physical setups~\cite{Xu:10-1,Xu:10-2,Auccaise:11}, which have 
demonstrated that such phenomena are quite general in real conditions. Remarkably, a recent investigation~\cite{Silva:12}  
experimentally verified the single sudden change and the correlation freezing for the negativity of quantumness 
in Bell-diagonal states, which turns out to be equivalent to the 1-norm GQD. Therefore, further experiments 
aiming at those properties and also at double sudden changes could greatly benefit from the  
correlation dynamics derived in this work. Moreover, we have found a surprising connection between the 
sudden change behavior and the infinite-order quantum phase transition in the XXZ model. Such kind of connections 
appear as a promising direction for investigating quantum criticality. We leave these 
possibilities as future challenges, as well as generalizations for more general states and for non-Markovian dynamics.

\begin{acknowledgments}

This work is supported by the Brazilian agencies CNPq, CAPES, FAPERJ, and the Brazilian National Institute for 
Science and Technology of Quantum Information (INCT-IQ).

\end{acknowledgments}


\end{document}